\documentclass{osa-article}

\journal{osajournal}
\setprjcopyright

\articletype{Research Article}

\begin{document}

\title{Observation of gain-pinned dissipative solitons in a microcavity laser}

\author{Maciej Pieczarka,\authormark{1,2,*} Dario Poletti,\authormark{3,4} Christian Schneider,\authormark{5} Sven H\"{o}fling,\authormark{5,6} Elena A. Ostrovskaya,\authormark{2}
Grzegorz S\k{e}k,\authormark{1} and Marcin Syperek\authormark{1}}

\address{\authormark{1}Department of Experimental Physics, Wroc\l{}aw University of Science and Technology, Wyb.~Wyspia\'nskiego 27, 50-370 Wroc\l{}aw, Poland\\
\authormark{2}ARC Centre of Excellence in Future Low-Energy Electronics Technologies and Nonlinear Physics Centre, Research School of Physics, The~Australian National University, Canberra, ACT 2601, Australia\\
\authormark{3}Science and Math Cluster, Singapore University of Technology and Design, 8 Somapah Road, 487372 Singapore\\
\authormark{4}EPD Pillar, Singapore University of Technology and Design, 8 Somapah Road, 487372 Singapore\\
\authormark{5}Technische Physik and Wilhelm-Conrad-R\"{o}ntgen-Research Center for Complex Material Systems, Universit\"{a}t W\"{u}rzburg, Am Hubland, 97074, Würzburg, Germany\\
\authormark{6}SUPA, School of Physics and Astronomy, University of St. Andrews, St. Andrews, KY 16 9SS, United Kingdom
}

\email{\authormark{*}maciej.pieczarka@pwr.edu.pl} 


\begin{abstract*}
We demonstrate an experimental approach to create dissipative solitons in a microcavity laser. In particular,  we shape the spatial gain profile of a quasi-one-dimensional microcavity laser with a nonresonant, pulsed optical pump to create spatially localised structures, called gain-pinned dissipative solitons that exist due to the balance of gain and nonlinear losses and are confined to a diffraction-limited volume. The ultrafast formation dynamics and decay of the gain-pinned solitons are probed directly, showing that they are created on a picosecond timescale, orders of magnitude faster than laser cavity solitons. All of the experimentally observed features and dynamics are reconstructed by using a standard complex Ginzburg-Landau model.
\end{abstract*}

\section{Introduction}
\begin{figure*}[ht]
\centering
\includegraphics[width=\textwidth]{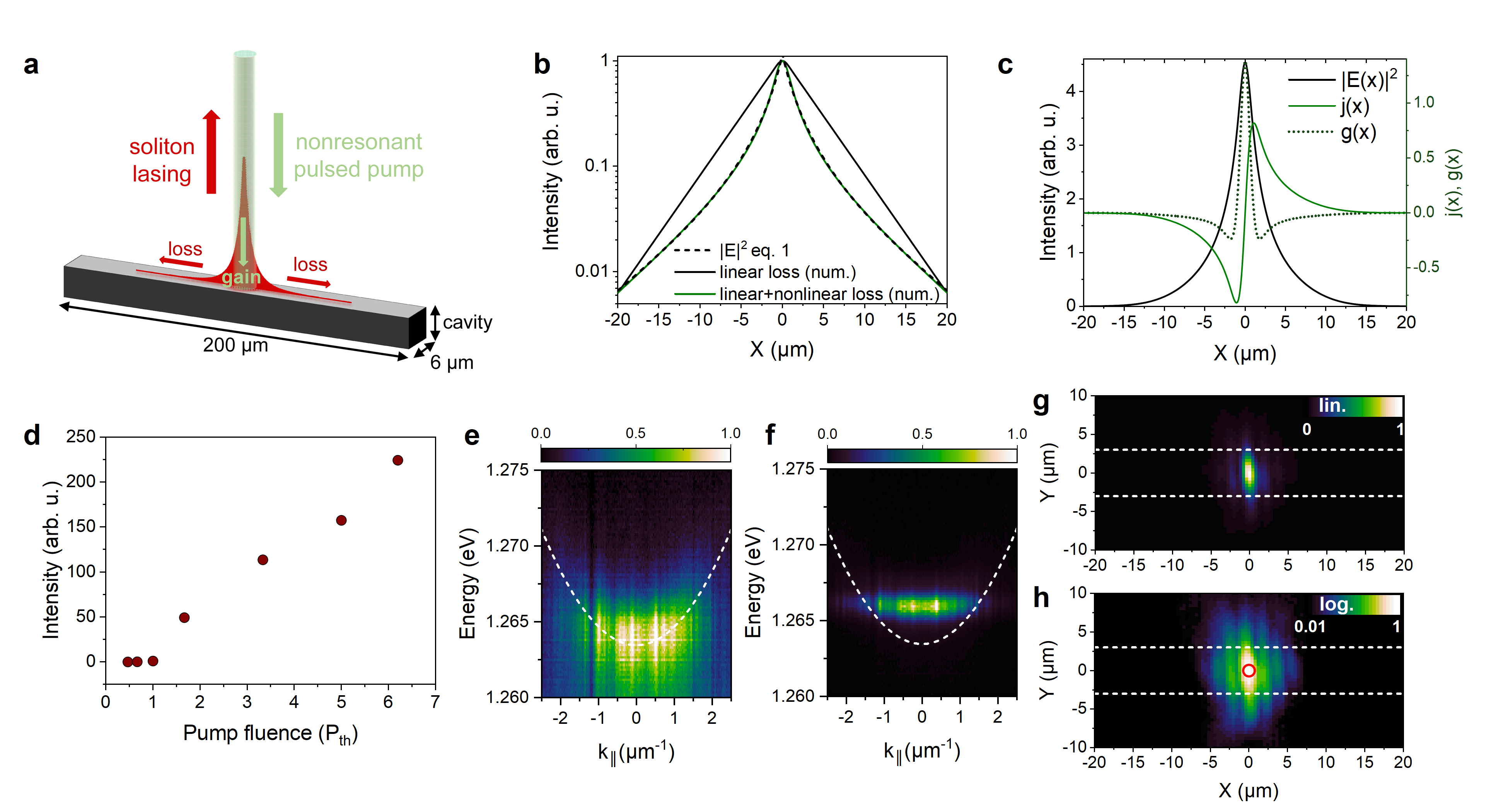}
\caption{\textbf{Gain-pinned dissipative soliton and the experimental characterisation of the one-dimensional microcavity laser.} (a)~Schematics of the experimental realisation. The pumping laser beam is focused on a small area of the quasi-one-dimensional microcavity laser supporting a pinned soliton mode. The dimensions of the microlaser stripe are indicated in the figure. (b) Dissipative soliton shapes calculated numerically with a constant linear pump of finite width under assumption of linear (black line) and nonlinear loss (green line) in the system (parameters are given in Theoretical model details section) and the analytical expression from eq.~\ref{eq:soliton} is plotted with a dashed line. (c) Dissipative soliton profile $|E(x)|^2$ plotted together with the local energy flux $j(x)$ and the local energy generation $g(x)$ characterising the gain-loss distribution of the mode. (d) Power dependent input-output series of an investigated microcavity laser showing typical threshold behaviour. (e,f) Far-field spectra of the lasing mode (e) below, and (f) at the lasing threshold. Blueshift and narrowing of the lasing mode is observed. Dashed line indicates the cavity photon momentum dispersion. (g) Real-space image of the spatial shape of the lasing soliton mode for $P=3.3P_{th}$. (h) The same as in (g) but in logarithmic colour scale to enhance visibility of the low intensity signal. Red circle indicates the gain spot.}
\label{fig:Fig1}
\end{figure*}

The last decades of research in nonlinear optics have uncovered an immense wealth of systems and material configurations in which solitons can be created \cite{Akhmediev2008}. In conservative optical systems, temporal or spatial solitons are supported by nonlinearity compensating for the dispersion or diffraction of light in the propagating geometry \cite{Chen2012,Eisenberg1998,Aitchison1992}. Real-world photonic devices suffer from intrinsic losses, e.g. via photon escape out of the structure, and it is essential to achieve the balance not only between the dispersion or diffraction and nonlinearity, but also in the energy flow, i.e. between gain and loss of the system, to support self-sustaining solitary modes or localised structures \cite{Coullet2002}. Temporal dissipative Kerr solitons, recently realised in ring microcavity resonators, are considered to be a very promising platform for various applications in miniaturisation of time standards, frequency metrology systems \cite{Yi2015,Herr2014} or mode-locked lasers \cite{Grelu2012}. Spatially localised dissipative structures, called cavity solitons \cite{Maggipinto2000}, have been successfully created in broad area vertical cavity surface emitting lasers (VCSELs) \cite{Barland2002,Genevet2008,Gustave2017}. In this configuration, the device is kept below the lasing threshold, while the use of an additional external coherent holding laser beam, coupled to an external cavity mirror, or to a saturable absorber to set up an optical bistability condition, leads to the creation of stable localised modes. Their control is implemented with an additional writing laser beam or pulse \cite{Pedaci2008,Elsass2010}. The spatially localised cavity solitons are not created in a propagating geometry in the device, but are confined within the optical microcavity containing the active medium and can move in the transverse (perpendicular to the cavity) direction.

In the opposite regime, when the laser device is driven above the threshold, the spatially uniform gain alone \cite{Hachair2006} (without a coherent holding beam) is not capable of sustaining a localised bright mode which becomes unstable due to the action of the gain outside of the soliton core. Therefore, another approach has to be implemented to overcome the competition between the coherent holding beam and the gain of the lasing field. One can imagine modifying the spatial gain profile in the device to tackle this issue. A straightforward way is to contain gain in a small spatial volume, where the loss outside of this \emph{hot spot} can provide a balance between gain and loss around the bright soliton core \cite{Zezyulin2011}, see Fig.~\ref{fig:Fig1}a and Fig.~\ref{fig:Fig1}c. Dissipative solitons pinned by a localised gain have been intensively studied theoretically, and their realisations in various systems were proposed \cite{Malomed2014}. These localised structures, known as gain-pinned solitons, have been predicted to be robust and stable over a wide range of parameters even in the absence of Kerr nonlinearity in the medium. In particular, a one-dimensional complex Ginzburg-Landau model with an infinitesimally localised gain in a dissipative medium supports an exact dissipative soliton solution \cite{Malomed2014,Lam2009}:
\begin{equation}
    E(x) = A{\sinh(\kappa(|x|+\xi))}^{(-1+i\mu)}
\label{eq:soliton}
\end{equation}
where $A$ is the amplitude of the mode, $\mu$ is the chirp coefficient and $\kappa$, $\xi$ determine the shape of the soliton envelope function. It is not a generic solution, as the analytical form is only available under a constraint on the parameters. Nevertheless, it has been shown numerically that this type of solution represents a broader family of stable localised modes \cite{Lam2009} and the gain-pinned dissipative soliton solution, eq.~\ref{eq:soliton}, is an attractor \cite{Lam2009,Malomed2014} of the complex Ginzburg-Landau model (see Theoretical model details) and is independent of initial conditions. Moreover, the localised dissipative solitons are expected to be stabilised in systems with non-negligible nonlinear losses \cite{Malomed2014,Lam2009}  or in absence of nonlinear losses in a system with cubic gain \cite{Borovkova2012}.

In a realistic model of an experiment, the gain spot has a finite spatial width convolved with the analytical shape of the dissipative soliton, see Fig.~ \ref{fig:Fig1}b. In the simplest case of linear losses in the cavity (photon escape rate), without the bulk Kerr nonlinearity, the mode profile is trivial and is characterised by exponential decay outside of the gain spot (black line in Fig.~\ref{fig:Fig1}b) \cite{Zezyulin2011,Malomed2014}. However, the shape is considerably changed when nonlinear losses, i.e. two-photon absorption, are also present (green line in Fig. \ref{fig:Fig1}b and dashed line showing the analytical fit $|E(x)|^2$ of eq.~\ref{eq:soliton}). In this case, the dissipative nonlinearity focuses the robust localised structure to the volume limited by the gain spot, which is translated into the modified shape.

Contrary to solitons in conservative systems, dissipative solitons depend on the external energy supply and their existence is dictated by the total balance between gain and loss \cite{Akhmediev2008,Malomed2014}. In a realistic model with finite gain, the spatial profile of a gain-pinned dissipative soliton (eq.~\ref{eq:soliton}) can be expressed as a complex field $E(x)=A(x)e^{i\varphi(x)}$ with the local amplitude $A(x)$ and the phase $\varphi(x)$. To verify the gain and loss balance, one can calculate the local energy flux $j(x)= A^2(x) \frac{d \varphi(x)}{d x}$ and its derivative $g(x)=\frac{d j(x)}{d x}$, which is the measure of the local gain $g(x)>0$ or loss $g(x)<0$. The numerically calculated gain-pinned soliton profile $|E(x)|^2$ and $j(x)$, $g(x)$ are presented in Fig.~\ref{fig:Fig1}c. The dissipative soliton is associated with the typical spatial profiles of the abovementioned quantities. Namely, the energy flux $j(x)$ is zero at the soliton centre, negative on the left and positive on the right, reflecting the fact that the energy flows from the centre to the tails of the mode, with the sign of the flux indicating the flow direction. The local energy gain $g(x)$ is positive in the centre, where the gain spot is located, and negative outside, indicating energy loss. The gain-loss balance criterion can be expressed as
\begin{equation}
    \int^{+\infty}_{-\infty} g(x) dx = 0,
\label{eq:balance}
\end{equation}
which is fulfilled for the gain-pinned dissipative soliton \cite{Akhmediev2008,Lam2009,Malomed2014,Zezyulin2011}.

In this work, we present a \emph{proof-of-concept} experimental realisation of a gain-pinned spatial dissipative soliton in the experimental configuration schematically shown in Fig. \ref{fig:Fig1}a. In this case, these localised structures are a type of cavity solitons, as they are realised and confined within the cavity of the laser and are not propagating in the longitudinal direction. We create a gain spot in an empty cavity with a focused, pulsed and nonresonant laser beam tuned above the bandgap of the cavity material (GaAs), and generate a spatial soliton that decays in time. The observed lasing mode is confined to a volume around the area of the resolution-limited gain. Subsequently, we track the ultrafast dynamics of the soliton formation, the evolution of the spatial profile and the profile narrowing, as well as the far-field emission by streak camera measurements. Additionally, we measure the soliton onset times of the order of few picoseconds, which is orders of magnitude faster than the previously reported cavity soliton manipulation dynamics \cite{Gustave2017,Pedaci2008,Elsass2010,Hachair2005}, and is comparable to bright exciton-polariton soliton excitation timescales \cite{Sich2012}. We successfully reconstruct all observed dynamical features by employing a numerical complex Giznburg-Landau model.

\section{Experimental results}

\begin{figure*}[ht]
\centering
\includegraphics[width=\textwidth]{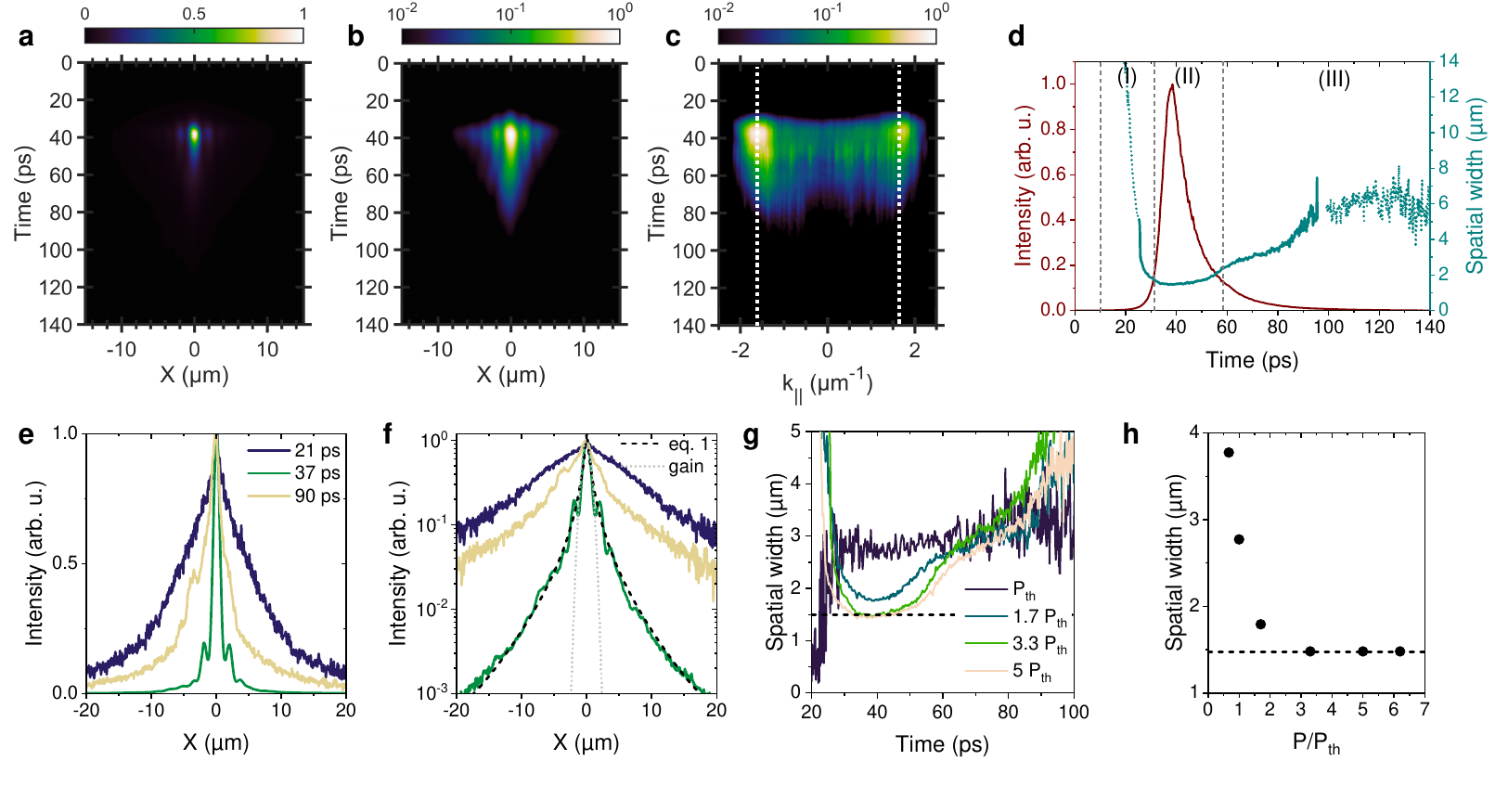}
\caption{\textbf{Experimental dynamics of the soliton creation and decay.} (a,b) Time-resolved spatial dynamics of the dissipative soliton emission in (a) linear and (b) logarithmic colour scale, respectively, at pumping power $P=3.3~P_{th}$. (c) Far-field dynamics of the emission, where two distinct wavevectors are visible and indicated with white dotted lines. (d) Time dependency of the dissipative soliton signal intensity and its spatial width (full width at half maximum - FWHM). Three different stages in the time dynamics are indicated in the figure: (I) lasing onset, (II) soliton creation and (III) decay and spreading at later times. Region with low signal-to-noise ratio of the spatial width data is depicted with dotted line. (e,f) Spatial distributions of the emission at different times indicated in the figure. The dashed curve is the dissipative soliton envelope curve (eq.~\ref{eq:soliton}) fitted to the experimental data $|E|^2$  and the black dotted line indicates the Gaussian laser gain spot. (g) Dissipative soliton compression dynamics under different pumping powers showing decrease of the spatial width with increasing number of photons and carriers in the system. (h) Minimal spatial width of the soliton, showing density-dependent narrowing of the dissipative soliton. Dashed line in (g), (h) indicate the spatial resolution of the setup, simultaneously defining the minimum measurable width of the gain spot.}
\label{fig:Fig2}
\end{figure*}

Following the existing theoretical proposals, we implement an experimental realisation of one-dimensional spatial dissipative solitons in a semiconductor microcavity laser, as shown in Fig. \ref{fig:Fig1}a. The cavity structures were processed by lithography and etching from a planar GaAs-based VCSEL sample in the form of stripes of hundreds of micrometres in length and only a few in width, creating an effectively one-dimensional transverse confinement of optical modes (see Fig. \ref{fig:Fig1}a and Experiment details section for details). The localised gain is provided by a pulsed laser source, frequency-tuned above the GaAs bandgap and to one of the reflectivity minima of the cavity for the efficient photo-excitation of electron-hole pairs. The laser spot is focused to a diffraction limited spot on the surface of the cavity stripe. In contrast to typical GaAs-based waveguide designs for optical soliton experiments, where the detrimental effects of nonlinear losses are kept to a minimum \cite{Eisenberg1998,Aitchison1992}, our sample design places the fundamental cavity mode in the spectral range where the refractive index nonlinearity is small and defocusing and, most importantly, where significant nonlinear losses due to the two-photon absorption are expected to occur \cite{Sheik-Bahae1990}. These material characteristics provide the conditions for the generation of gain-pinned dissipative solitons, as indicated by the theoretical models \cite{Malomed2014}. Additionally, the nonlinear loss channel is known to be more pronounced in photonic devices due to a small volume of the confined photon mode \cite{Combrie2008,Notomi2008,Yuce2012,Soljacic2004}, effectively lowering the total power densities at which the nonlinear effects occur. This property makes our system most suitable for exploiting the nonlinear loss in the process of dissipative soliton formation.

The power dependent measurements of the output intensity of the device luminescence reveal a typical lasing threshold (where the lasing threshold value $P_{th}=29~pJ/\mu m^2$), see Fig. \ref{fig:Fig1}d, being accompanied by the emission linewidth narrowing and the distinct blueshift of the emission energy (by about 3 meV), as seen in the far-field (wave vector space) spectra in Figs. \ref{fig:Fig1}e and \ref{fig:Fig1}f. This energy shift originates from the local cavity refractive index change due to the free carriers generated in the GaAs spacer of the cavity volume by a nonresonant pump pulse \cite{Yuce2012,Said1992}. Subsequently, carriers relax to the quantum wells, creating an electron-hole plasma eventually providing gain for the lasing and the soliton pinning. Nevertheless, the positive energy shift does not influence the stability of the observed soliton creation, as the soliton dynamics are weakly affected when the local potential modification is absent (not shown). The broad momentum range of the far-field spectrum in Figs. \ref{fig:Fig1}e and \ref{fig:Fig1}f reflects the strong localisation of the soliton observed in real space, Figs. \ref{fig:Fig1}g and \ref{fig:Fig1}h. The spatial extent of the soliton is constrained to the diffraction-limited gain spot ($\sim 1.5~\mu m$) in the direction along the long axis of the stripe, and the sample dimensions in the direction of the short axis (mode width is about $4~\mu m$). The mode volume could possibly be confined to an even smaller area by employing narrower microwire cavities or photonic crystal nanocavities \cite{Notomi2008}.

The pulsed excitation used in the experiment results in a non-stationary, decaying lasing of the dissipative soliton owing to the finite lifetime of the excited carriers and cavity photons in the system. Hence, we performed an analysis of the shape and dynamics of the solitary wave in the direct time-resolved experiment imaging the emission with a streak camera (see Experiment details section). The dynamics of the near and far-field emission patterns along the microcavity wire excited above the lasing threshold are presented in Figs. \ref{fig:Fig2}a-c with analysis of the spatial width and the signal intensity shown in Fig. \ref{fig:Fig2}d. One can distinguish three stages of the soliton laser dynamics. Firstly, the pump pulse creates high-energy electrons and holes in the barrier which relax and form a gain medium within the quantum well states. Then, after about $10~ps$, the lasing occurs with a rapid narrowing of the spatial width down to the optical resolution limit of the setup (approx. $1.5~\mu m$). Subsequently, the soliton pulse maintains its narrow width with the strongest emission of photons, and eventually it decays and spreads at later times (after $t > 60~ps$), which is evidenced in the increase of the spatial width, Fig. \ref{fig:Fig2}d. This latter stage is the linear regime, where the gain and nonlinearity are weak and cannot sustain the soliton shape. The duration of the solitary lasing is short, about $20-30~ps$, with an ultra-short rise time of approximately $3~ps$, being limited by the temporal resolution of the setup. The observed fast response is a result of the onset of the stimulated emission and the ultrafast dynamics of electronic semiconductor nonlinearities occurring on the timescales of a few picoseconds \cite{Yuce2012,Xie2016,Altug2006,Harding2007}.

The three characteristic regimes in the dynamics are also distinguishable in the spatial shape of the emission as presented in Figs. \ref{fig:Fig2}e, f. At the onset of lasing, when the emission intensity is low, the lasing mode is weakly localised around the gain spot with exponential spatial decay. Subsequently, the lasing intensity rises, reaching a maximum at around $t = 37~ps$, whereby the nonlinear trapping and formation of a strongly localized dissipative soliton occurs at high photon densities. The mode shape, see Fig. \ref{fig:Fig2}f, reveals the solitary solution discussed in Fig. \ref{fig:Fig1} and follows the analytical expression of Eq. \ref{eq:soliton}. The strong spatial focusing is most likely caused by the presence of nonlinear losses in the system, since any other nonlinearity (e.g. due to the Kerr effects, or free-carriers related) is defocusing and would lead to a trivial spatial shape, as shown in Fig. \ref{fig:Fig1}b. Additionally, we can safely rule out thermal effects, which have much longer timescales and manifest themselves in a redshift of the lasing mode energy \cite{Anguiano2019}.

The far-field spectrum of the dissipative soliton contains two distinct peaks (Fig. \ref{fig:Fig2}c) due to the outward propagation of photons from the gain spot. This shape is the consequence of a dynamical balance of the energy flow characteristic of dissipative solitons, as discussed in the introduction: an inflow due to the nonresonant pump and outflow because of the linear and nonlinear losses, which stabilises the soliton \cite{Akhmediev2008,Zezyulin2011,Malomed2014} (see the scheme in Fig. \ref{fig:Fig1}a and the shape analysis in Fig.~\ref{fig:Fig1}c). Moreover, backscattering of these propagating waves on the intrinsic disorder of the sample causes the characteristic interference pattern observed on top of the solitary mode. This effect is seen both in time integrated, Fig. \ref{fig:Fig1}h, and time-resolved images, in Fig. \ref{fig:Fig2}b. The disorder scattering in the sample adds a small spatial modulation to the soliton envelope, the latter being independent of the particular position on the sample. At longer times after the pulse, the nonlinear trapping weakens along with the lasing signal and the mode is once again exponentially confined with additional disorder-induced modulation. The nonlinear mechanism of the mode trapping is also manifested in the density-dependent measurements, where one can see sharp narrowing of the soliton width above the lasing threshold, Fig. \ref{fig:Fig2}g, h. The minimum measured soliton width is due to the setup resolution, being also the limitation for the gain spot diameter, and is indicated by an horizontal dashed line in Fig. \ref{fig:Fig2}g, h. We note that no localisation was observed when the structure was pumped with a large pump spot of about 40 $\mu m$, which rules out the localisation caused by spatial hole burning effect of trivial localisation by the local disorder.

\section{Numerical modelling results}

To verify the interpretation of our experiment proposed above and understand the observed dynamics, we perform numerical simulations based on a complex Ginzburg-Landau equation including nonresonant pumping and dissipation in the system, taking into account nonlinear losses and simplified dynamics of the photo-excited reservoir of carriers (see equations \ref{eq:model1} and \ref{eq:model2} in Theoretical model details section). The reservoir of carriers not only provides the gain to the system, but also leads to a local modulation of the cavity refractive index, so-called linewidth enhancement factor, which drops down when the reservoir is depleted by the stimulated laser emission and nonradiative decay. 

\begin{figure}[ht]
\centering
\includegraphics[width=0.75\textwidth]{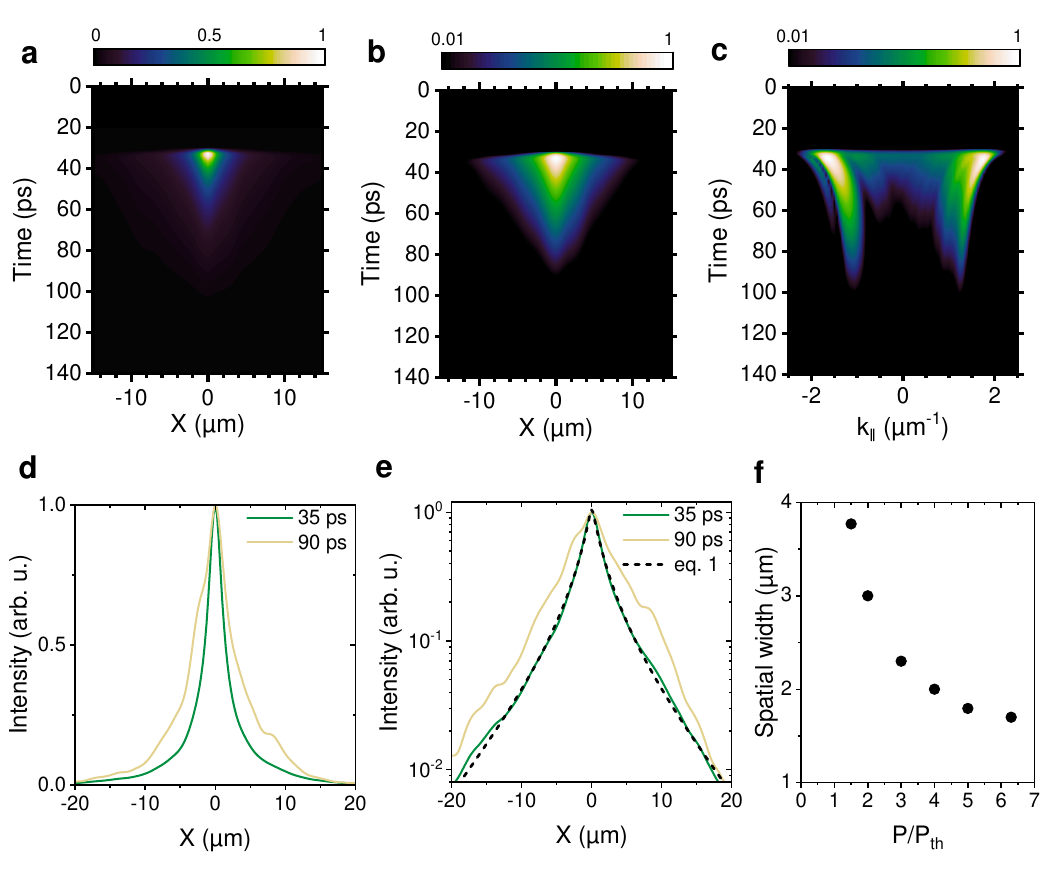}
\caption{\textbf{Numerical simulations of the soliton dynamics.} Numerical simulations of the recorded time dynamics presented in Figure \ref{fig:Fig2}. (a,b) Spatial dynamics of the computed soliton mode emission in (a) linear and (b) logarithmic colour scale. All characteristics of experimental data are well reproduced. (c) Far-field dynamics of the mode emission. (d,e) Spatial distributions of the lasing mode at two different times: presenting the soliton mode at $35~ps$ and the diffracted mode at a later time $90~ps$, with the intensity plotted on (d) linear and (e) logarithmic scale. The roughness of the profiles is due to the presence of disorder similar to the experiments. Analytical fit to the soliton profile at $35~ps$ is drawn with a dashed line. (f) Narrowing of the spatial width of the lasing modewith pump amplitude, demonstrating the nonlinear narrowing above a pump threshold. The mode width is limited by the width of the simulated gain spot.}
\label{fig:Fig3}
\end{figure}

Our numerical simulations successfully reconstruct, as shown in Fig.3, all the characteristic features seen in the experiment: the soliton shape and dynamics in the real space as well as the far-field spectra.  As shown in Fig. 3c, the soliton is localised around the gain spot in real space, and its Fourier spectrum contains two main wavevector components. Power-dependent simulations also yield the narrowing of the spatial width, see Fig. 3f, which is the manifestation of the nonlinear trapping mechanism. Inclusion of nonlinear losses in the model is essential for reproducing the experimental shape of the soliton, as the model with only linear losses captures neither the dynamics presented in Fig.~\ref{fig:Fig3}, nor the width narrowing shown in Fig.~\ref{fig:Fig1}b. Our model also includes a random disorder potential with similar characteristics to the one measured experimentally, which slightly modulates the soliton shape. The soliton shape is found to be robust to the change of the particular realisations of the disorder, which are kept within the experimentally measured values.

\section{Conclusions}

We have demonstrated one-dimensional gain-pinned dissipative solitons in VCSELs for which: (i) the experimental realization is particularly convenient due to a nonresonant pumping, (ii) the  soliton size is limited by the gain profile and is therefore much smaller than the typical VCSEL solitons \cite{Barland2002,Genevet2008,Gustave2017,Pedaci2008} and comparable to the size of exciton-polariton solitons \cite{Sich2012,Skryabin2017}, and (iii) the soliton formation dynamics, being driven by stimulated laser emission, is an ultra-fast process in the range of single picoseconds, orders of magnitude faster than cavity solitons \cite{Hachair2005,Elsass2010} and of the same order of magnitude as bright exciton-polariton solitons \cite{Sich2012,Tanese2013}. The temporal decay of the lasing signal observed in our experiment is due to the pulsed regime of excitation, whereby the gain-pinned dissipative soliton is created in a "single-shot" regime. Continuous wave (CW) pumping would be capable of maintaining a steady-state soliton lasing with a CW gain. This offers a possibility to create a stable, steady-state dissipative soliton localised in a device.

Although our study explores a quasi-one-dimensional version of a VCSEL laser, our results pave the way towards creating stable two-dimensional solitary modes, which, as predicted theoretically \cite{Kartashov2011}, are within reach in modern semiconductor microcavities of similar design. Manipulation of the soliton position and its transversal propagation can be performed by spatial modulation of the excitation beam, using spatial light modulators (SLMs), offering exciting possibilities for the creation of two-dimensional vortex solitons \cite{Kartashov2011,Ma2017,Lobanov2011,Huang2013} or multi-soliton structures \cite{HouTsang2011,Vladimirov2002}. Furthermore, gain-pinned solitons presented here, due to their robustness and simple realisation, could be arranged in lattices forming a platform for simulations of classical Hamiltonians \cite{Berloff2017,Nixon2013}, studies of complex topological ordering \cite{Tosi2012,Pal2017}, and spontaneous symmetry breaking in laser systems \cite{HouTsang2011,Kartashov02011,Hamel2015}.

\section*{Experiment details}
The sample under investigation is an AlAs/GaAs $\lambda/2$-long planar microcavity composed of two distributed Bragg reflectors (DBRs) enclosing two stacks of four InGaAs/GaAs quantum wells located at the antinodes of the photon field. The ground state of the quantum well is located around $1.262~eV$. The cavity resonance is placed in the spectral region, where refractive index nonlinearity is minimized and defocusing and significant nonlinear losses due to the two-photon absorption are expected to occur (calculations are made based on cavity spacer GaAs parameters\cite{Sheik-Bahae1990}).  The one-dimensional microwires were created via electron beam lithography and etched using electron cyclotron-resonance reactive-ion-etching. The semiconductor-air interface on sidewalls of the microwire provides the spatial confinement of the photon modes in one of the in-plane directions. During the experiment, the sample was kept in a continuous flow liquid helium cryostat at a temperature $T= 5~K$. 

The excitation is provided by a mode-locked tunable Ti:Sapphire laser emitting $140~fs$ pulses with $76~MHz$ repetition. The laser spot was focused via a high numerical aperture objective lens (NA = 0.42) to a diffraction limited spot of about $1.5~\mu m$ width. For the efficient injection of free carriers into the QWs, the laser wavelength was tuned to the high-energy reflectivity minima of the microcavity, located at around $1.55~eV$, for the above GaAs bandgap excitation, and the second around $1.47~eV$. Excitation above the cavity spacer material bandgap provides a local modulation of the cavity refractive index due to photogenerated excessive carriers. The emission from the sample was collected via the same microscope objective, and then further transferred through a set of achromatic lenses to a spectrometer for near-field and far-field imaging. The monochromator ($0.5~m$ focal length, Princeton Instruments) outputs were connected to a two-dimensional InGaAs near-infrared camera (NIRvana Princeton Instruments) and to a Hamamatsu streak camera (temporal resolution of about $3~ps$). For imaging purposes, the monochromator grating was set to the zero-order mode.

\section*{Theoretical model details}
The soliton dynamics is modelled by a complex one-dimensional Ginzburg-Landau equation describing the electric field envelope $E(x,t)$ of the lasing mode coupled to a rate equation for the density of the carrier reservoir $N(x,t)$, i.e. the gain medium:
\begin{equation}
\begin{split}
    &\frac{\partial}{\partial t}E(x,t) = \frac{ic^2}{2k_{c}n^2_c} \frac{\partial^2}{\partial x^2}E(x,t)+
   \\ &+\frac{1}{2} \left( \Gamma N(x,t) - \gamma_c - \beta |E(x,t)|^2 \right)E(x,t) - \\ &-iV(x)  E(x,t) - i\alpha N(x,t)E(x,t),
\end{split}
\label{eq:model1}
\end{equation}

\begin{equation}
\frac{\partial}{\partial t} N(x,t) = P(x,t) - \gamma N(x,t) - \Gamma |E(x,t)|^2 N(x,t).
\label{eq:model2}
\end{equation}

Here, the cavity photons are described by an effective mass along the microcavity stripe $m^{*}=\frac{E_c}{(n_c ⁄ c)^2}$, where $E_c = \hbar k_c c$ is the cavity photon energy at $k_\parallel = 0$, $n_c$ is the cavity refractive index, $c$ is the speed of light, and $k_c$ is the confined longitudinal mode wavenumber. The gain in the system is described by the coefficient $\Gamma$, and the linear loss (cavity photon lifetime) is denoted by $\gamma_c$. The nonlinear losses (e.g., two-photon absorption) in the system are denoted by $\beta$. The reservoir decay rate is determined by $\gamma$. The nonresonant pumping of the system is described by the term $P(x,t)$. This term can be constant for continuous wave simulations or can be expressed as $P(x)\delta(t=0)$ for a pulsed excitation, setting the initial spatial density distribution of carriers in the microcavity. Local modifications of the cavity refractive index are introduced through the carrier density and scaled with the linewidth enhancement factor parameter $\alpha$. The material disorder is described as a static potential $V(x)$.

In the case of CW excitation, the model can be reduced to a single equation under the assumption of the carrier density $N(x,t)$ adiabatically following the field evolution $|E(x,t)|^2$. The equation \ref{eq:model2} can be solved for a CW pump $P(x,t)=P(x)$:
\begin{equation}
    N(x,t) = \frac{P(x)}{\gamma + \Gamma |E(x,t)|^2}.
\end{equation}

Simulations were performed with a following set of experimentally valid parameters: $\gamma_c=1~ps^{-1}$, $\gamma=1~ns^{-1}$, $m^*=3.12\cdot10^{-5} m_0$, where $m_0$ is the free electron mass, $\Gamma=0.01~\mu m/ps$,  $\beta= 1~\mu m/ps$ and $\alpha=4.6 \cdot 10^{-3} ~\mu m/ps$.

\section*{Funding}
The work was supported by the National Science Centre in Poland, by grant No. 2016/23/N/ST3/01350, and by the Polish National Agency for Academic Exchange. The W\"{u}rzburg group gratefully acknowledges support by the State of Bavaria. The work at the Australian National University was supported by the Australian Research Council.

\section*{Acknowledgments}
M.P. would like to acknowledge fruitful discussions with Yuri S. Kivshar and Dragomir Neshev. Assistance by Fabian Langer, Monika Emmerling, and Adriana Wolf during sample fabrication is acknowledged.

\section*{Disclosures}
The authors declare no conflicts of interest.

\bibliography{references}
\end{document}